\documentclass[epj]{svjour}
\usepackage{amsmath,amssymb}
\usepackage{times}
\usepackage{txfonts}
\usepackage[pdftex]{graphicx,color}
\usepackage{dcolumn}
\usepackage{bm}
\usepackage[mathlines]{lineno}

\usepackage{lscape}

\everymath{\displaystyle}

\renewcommand{\a}{$\alpha$}
\newcommand{\natPt}{$^\mathrm{nat}$Pt}
\newcommand{\mPt}{$^{195m}$Pt }

\begin{document}
\titlerunning{
Influence of secondary neutrons on alpha-particle induced reaction cross section ...
}
\title{
Influence of secondary neutrons on alpha-particle induced reaction cross section measurement below the Coulomb barrier
}

\authorrunning{Yamato Fujii, Naohiko Otuka et al.}
\author{
Yamato Fujii\inst{1.2}\thanks{fujii24@qr.see.eng.osaka-u.ac.jp (corresponding author)}
\and
Naohiko Otuka\inst{1,3}\thanks{n.otsuka@iaea.org (corresponding author)}
\and
Kenta Sugihara\inst{3,4,5}
\and
Masayuki Aikawa\inst{6,7,8}
\and
Hiromitsu Haba\inst{3}
\and
Isao Murata\inst{2}
}
\institute{
Nuclear Data Section, Division of Physical and Chemical Sciences, Department of Nuclear Sciences and Applications, International Atomic Energy Agency, A-1400 Wien, Austria
\and
Graduate School of Engineering, The University of Osaka, Suita 565-0871, Japan
\and
Nishina Center for Accelerator-Based Science, RIKEN, Wako 351-0198, Japan
\and
High Energy Accelerator Research Organization (KEK), Oho, Tsukuba, Ibaraki 305-0801, Japan
\and
The Graduate University for Advanced Studies (SOKENDAI), Hayama, Kanagawa 240-0193, Japan
\and
Faculty of Science, Hokkaido University, Sapporo 060-0810, Japan
\and
Graduate School of Biomedical Science and Engineering, Hokkaido University, Sapporo 060-8638, Japan
\and
Global Center for Biomedical Science and Engineering, Faculty of Medicine, Hokkaido University, Sapporo 060-8648, Japan
}

\date{Received: date / Revised version: date}

\abstract{
The influence of the secondary neutrons on measurements of alpha-particle activation cross sections below the Coulomb barrier was studied for the \natPt(\a,x)\mPt reaction.
We characterized the secondary neutron field by using the particle transport simulation code PHITS,
and estimated the \natPt(n,x)\mPt yields by using the characterized neutron spectra and the \natPt(n,x)\mPt cross sections in the JENDL-5/A library.
We confirmed that the unexpectedly high \natPt(\a,x)\mPt cross sections below the Coulomb barrier measured by us are explained well by the \natPt(n,x)\mPt reaction induced by the secondary neutrons. 
This indicates that the secondary neutron effect is sometimes not negligible even in low energy charged-particle activation cross section measurements.
We also studied the influence of the secondary light charged particles by the same approach,
and confirmed that their influence is negligible.
}

\maketitle
\onecolumn
\section{Introduction}
\label{sec:introduction}
Secondary particles are the outgoing particles produced by an interaction of a target nucleus with a beam particle.
They may undergo an interaction with another target nucleus (secondary reaction).
The cross section measured by irradiation of a relatively thick sample must be corrected for this effect.
In low energy neutron capture cross section measurements,
capture of a neutron scattered in the sample is known as the multiple scattering effect,
and correction for this effect is routinely done~\cite{Schillebeeckx2012}.

An interesting example in high energy charged-particle induced reaction measurements is seen in an article reporting on the $^{209}$Bi(p,x)$^{211}$At  production cross section at 200~MeV~\cite{Clark1982}.
Due to mass conservation,
this production is possible only by the secondary reaction,
namely $^{209}$Bi(\a,2n)$^{211}$At reaction following $^{209}$Bi(p,\a+x) reaction.
They studied the importance of the secondary reaction by performing measurements by varying the sample thickness.

A method to correct high energy charged-particle induced reaction cross sections for the secondary reaction effect was developed by the Hannover group~\cite{Michel1995,Luepke1993}.
They realized that the proton-induced reaction cross sections between 800 and 2600~MeV measured by them for many years suffered from the secondary reaction effect for a wide range of target nuclides (O to Au).
They developed a method to correct their cross sections for the effect caused by secondary neutrons and protons by using the high energy particle transport code HET~\cite{Armstrong1972},
and found that the corrections may exceed 90\% for reactions leaving product nuclides having mass close to the target nuclide mass such as (p,n) reactions.

In measurements of low energy charged-particle activation cross sections employing the stacked foil method,
extra foils are sometimes added at the end of the foil stack where the beam particles cannot reach due to the stopping power (i.e., outside of the range),
and presence of the secondary particle effect is checked by measuring the activities of these extra foils (e.g.,~\cite{Khandaker2013}).
It is, however, not a common practice to estimate the correction factor quantitatively for the secondary reaction effect in low energy charged-particle activation measurements.

We have recently measured activation cross sections for irradiation of platinum by alpha-particle beams at 30 and 50~MeV extracted from the RIKEN AVF cyclotron by the stacked foil method~\cite{Otuka2024,Otuka2025}.
Among the product nuclides investigated in these measurements,
we obtained unexpectedly high \mPt production cross sections in the low energy region.
The \natPt(\a,x)\mPt production threshold energy estimated by the nuclear mass table is below 1~MeV because this metastable state ($E_x\sim$260~keV) can be produced by $^{195}$Pt(\a,\a')\mPt inelastic scattering.
We expect that a measured excitation function exhibits influence of the Coulomb barrier between the alpha particle and platinum ($\sim$23~MeV).
However, the \mPt production cross sections measured by us show weak energy dependence over the subbarrier region.
Furthermore, we observed that the production cross sections measured with 50~MeV alpha-particle beams are systematically higher than those measured with 30~MeV alpha-particle beams below the Coulomb barrier.
As \mPt may be also produced by the $^{195}$Pt(n,n')\mPt reaction,
presence of the \natPt(\a,n+x) secondary neutrons may contribute to the measured \mPt yield. 

The 30~MeV alpha-particle beam extracted from the RIKEN AVF cyclotron is also utilized for production of $^{211}$At for medical applications through the $^{209}$Bi(\a,2n)$^{211}$At reaction.
To study the radiation shielding against the secondary neutrons,
the double differential thick target neutron yield was measured by using the time-of-flight method~\cite{Sugihara2020},
and it was reproduced well by the intranuclear cascade model INCL~\cite{Boudard2013} implemented in the Monte Carlo particle transport simulation code PHITS (Particle and Heavy Ion Transport code System)~\cite{Sato2018}.
It was also demonstrated that the combination of the PHITS code with the JENDL-5 alpha-particle sublibrary~\cite{Iwamoto2023} can be applied to prediction of neutron spectra of (\a,n) neutron sources such as the AmBe source~\cite{Ogawa2025}.
As it was done for high energy proton activation cross sections by the Hannover group,
it should be possible to study the secondary neutron effect in the \natPt(\a,x)\mPt cross section measurement by using particle transport calculation.

The purpose of the present work is to study the secondary neutron contribution to the measured \natPt(\a,x)\mPt production cross sections by using the secondary neutron spectra estimated by particle transport calculation.
This article is organized as follows:
A brief summary of the cross section measurement (Section 2) is followed by the description on the methods of secondary neutron field characterization (Section 3) and estimation of the yield of $^{195m}$Pt induced by secondary neutrons (Section 4).
The calculated secondary neutron fields and $^{195m}$Pt yields are presented in Section 5 and discussed in Section 6.
Finally, this work is summarized with key conclusions in Section 7.

\section{Cross section measurement}
\label{sec:measurement}
Pure metallic foils of natural platinum,
natural titanium and aluminium were stacked and inserted in a target holder made of aluminium alloy 5056 for irradiation by an alpha-particle beam extracted from the RIKEN AVF cyclotron.
The foil stack consisting of nine series of Pt-Al-Ti-Al followed by four additional Ti foils was irradiated at 29.0 MeV in 2023 (``30~MeV experiment")~\cite{Otuka2024},
and another foil stack consisting of nineteen series of Pt-Al-Ti-Ti was irradiated at 50.8 MeV in 2024 (``50~MeV experiment")~\cite{Otuka2025}.
Some parameters specifying the foil geometry  are summarized in Table~\ref{tab:sample}.
The alpha-particle beam energies are fully absorbed in these foil stacks.
The yield of \mPt (4.010 d) in each platinum foil was quantified by measuring 98.9 keV gamma-rays (11.7\%) by an HPGe detector.
The 98.9 keV gamma-rays are also emitted by $^{195}$Au (186.01 d),
which is formed directly by the \natPt(\a,x)$^{195}$Au reaction as well as decays of $^{195m}$Hg (41.6 h) and $^{195g}$Hg (10.53 h) following the \natPt(\a,x)$^{195m,g}$Hg reactions.
It is therefore hard to determine \mPt production cross section by measurement of the 98.9 keV gamma-rays only,
and we applied the simultaneous decay curve fitting method.
See more details about the experimental setup,
data reduction procedure and obtained cross sections in our previous articles~\cite{Otuka2024,Otuka2025}.

\begin{table}[hbtp]
\begin{center}
\caption{
Specification of foils irradiated at RIKEN
}
\label{tab:sample}
\begin{tabular}{lll}
\hline
Year of irradiation   & 2023           & 2024            \\
Reference             &\cite{Otuka2024}& \cite{Otuka2025}\\
\hline
Beam energy (MeV)     & 29.0           & 50.8            \\
Area (mm$^2$)         & 8$\times$8     & 8$\times$8      \\
Pt thickness ($\mu$m) & 5.796          & 10.103          \\
Ti thickness ($\mu$m) & 5.187          & 5.186           \\
Al thickness ($\mu$m) & 5.551          & 6.724           \\
\hline
\end{tabular}
\end{center}
\end{table}

\section{Neutron spectrum calculation}
\label{sec:calculation}

\subsection{INC and NDL options}
\label{sec:incndl}
We characterized the \natPt(\a,n+x) neutron spectrum in each platinum foil by PHITS ver.3.35.
PHITS is a general purpose Monte Carlo particle transport code simulating various physics processes including interactions of materials with particles (e.g., neutrons, protons, nuclei and mesons).
The code was validated by a comprehensive benchmark study~\cite{Iwamoto2017},
and applied to a wide range of applications such as radiation facility design, radiation therapy and protection, and space research.
Among various particle transport codes,
we selected PHITS for the present study because it demonstrates applicability to ($\alpha$,n) neutron field characterization~\cite{Ogawa2025},
allows use of the nuclear data libraries processed by users,
and offers access to the code package including source codes without registration fee.
For estimation of the (\a,n+x) reaction probability in a PHITS simulation,
one can adopt the (1) Intranuclear Cascade Model (INC) or (2) evaluated Nuclear Data Library (NDL).

The INC model describes a nuclear reaction as a superposition of particle motion and particle-particle collision.
It was originally developed for the description of high energy nuclear reactions including relativistic heavy-ion collisions (e.g.,~\cite{Nara1999}).
However, it has been known that the model often gives reasonable pictures even in the energy region below $\sim$100~MeV (e.g.,~\cite{Hayakawa1955,David2015}.
The Li\'ege Intranuclear Cascade model (INCL)~\cite{Boudard2013} is implemented in the PHITS code for the INC option.

Nuclear data libraries offer energy dependent cross sections predicted by reaction models such as the Hauser-Feshbach statistical model and adjusted to describe experimental cross sections,
and one may expect the performance of the NDL option is better than the INC option below $\sim$100~MeV.
However, the energy domain covered by the nuclear data libraries is typically up to 20~MeV and also very few libraries offer evaluated cross sections for charged-particle induced reactions.
Among the general purpose libraries,
the JENDL-5 library provides an alpha-particle sublibrary,
which is limited to interaction of an alpha-particle below 15~MeV with 18 light target nuclides.
Contrary,
the alpha-particle induced reaction cross sections calculated by TALYS-2.0~\cite{Koning2023} are compiled in the TENDL-2023 library~\cite{Koning2019} for all materials relevant to our current investigation up to 200~MeV.
We processed its ENDF-6 files by NJOY2016~\cite{MacFarlane2010},
and adopted the generated ACE files for the NDL option.

\subsection{Performance of INC and NDL options for $^{209}$Bi(\a,n+x) at 30~MeV}
\label{sec:Bi}
The performance of the INC option for characterization of neutron spectra has been already studied for the double differential thick target neutron yields from irradiation of a Bi sample by 30~MeV alpha-particles measured by the time-of-flight method at the RIKEN AVF cyclotron~\cite{Sugihara2020}.
They calculated the spectra by PHITS ver. 3.00 with the INC option,
and confirmed that it reproduces the experimental spectra.
However,
the NDL option was not examined in their study.
Therefore,
we estimated the $^{209}$Bi(\a,n+x) double differential thick target neutron yields with both INC and NDL options by using PHITS ver. 3.35  to determine which option is more suitable for evaluating the \natPt(\a,n+x) secondary neutron fluence in our present study.

\begin{figure}[hbtp]
\begin{center}
\includegraphics[width=0.6\textwidth]{"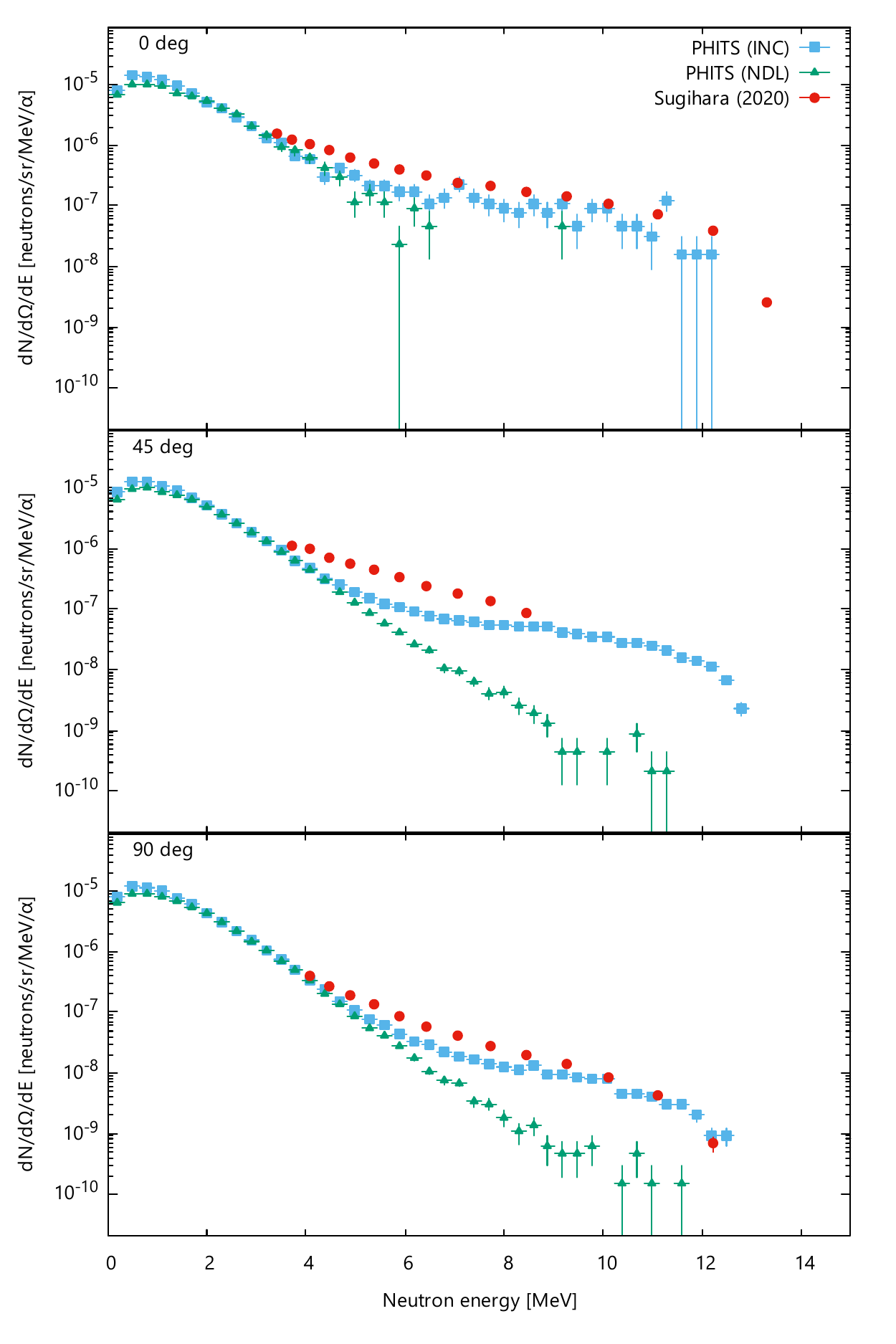"}
\caption{
Comparison of double differential thick target neutron yields for Bi irradiated by 30~MeV alpha-particles measured at RIKEN~\cite{Sugihara2020} with those predicted by PHITS INC and NDL calculations.
The error bars of the experimental yields are small and not visible. 
}
\label{fig:Bi}
\end{center}
\end{figure}

Figure~\ref{fig:Bi} shows the comparison of the measured and simulated neutron spectra.
We observe that both options give similar spectra in the low energy peak region around 1~MeV.
Both options underestimate the spectra in the preequilibrium region around 7~MeV,
and it is more significant in the NDL option.
In our study of secondary neutron effects in stacked foils,
the main contribution to the secondary neutron fluence comes from the neutrons around 1~MeV,
which are predominantly produced through compound nucleus processes and are expected to have an approximately isotropic angular distribution.
Since both INC and NDL options predict similar neutron yields in this energy region,
it is expected that the predicted production yields of \mPt from secondary neutrons will also be similar.
This understanding motivated use of both INC and NDL options in our evaluation of secondary neutron fields in the platinum irradiation experiments.

\section{Method for estimation of \mPt yield by secondary neutrons}
\label{sec:method}

\subsection{Estimation of secondary neutron spectra from \natPt(\a,n+x) reactions}
\label{sec:nfield}
To evaluate the contribution of the \natPt(n,x)\mPt reaction in the measured \natPt(\a,x)\mPt cross sections,
we estimated the  \natPt(\a,n+x) neutron fluence for all Pt foils (9 foils for 30~MeV irradiation and 19 foils for 50~MeV irradiations) using the PHITS code with both INC and NDL options.
We modelled the foils and target holder  but omitted other surrounding materials such as the beam collimator in the simulations.
The neutron fluence was obtained by the T-Track tally of the PHITS code.
To obtain the neutron fluence in a group-wise expression (neutrons/cm$^2$/\a) taking into account the resonance structure of the $^{194}$Pt(n,$\gamma$)\mPt cross section,
we divided the energy interval between 1.00$\times 10^{-11}$~MeV and 1.19433~MeV into 1632 groups adopting the ECCO-2000 group structure~\cite{Sartori1991} and above 1.20~MeV,
the energy bins were set at 10~keV intervals.
Our calculations,
performed with and without the target holder,
indicated its negligible influence for both the INC and NDL options.
Hereafter,
our discussion is based on PHITS calculations including the target holder.

\subsection{Renormalization of $^{194,195,196}$Pt(n,x)\mPt cross sections in JENDL-5}
\label{sec:leastsq}
In the JENDL-5 activation cross section library (JENDL-5/A~\cite{Iwamoto2022}),
three neutron-induced nuclear reactions $^{194}$Pt(n,$\gamma$)\mPt, $^{195}$Pt(n,n')\mPt, and $^{196}$Pt(n,2n)\mPt are responsible for production of \mPt from natural platinum.
The evaluated cross sections of these three reactions were used in our characterization of the secondary neutron field.

\begin{figure}[hbtp]
\begin{center}
\includegraphics[width=0.8\textwidth]{"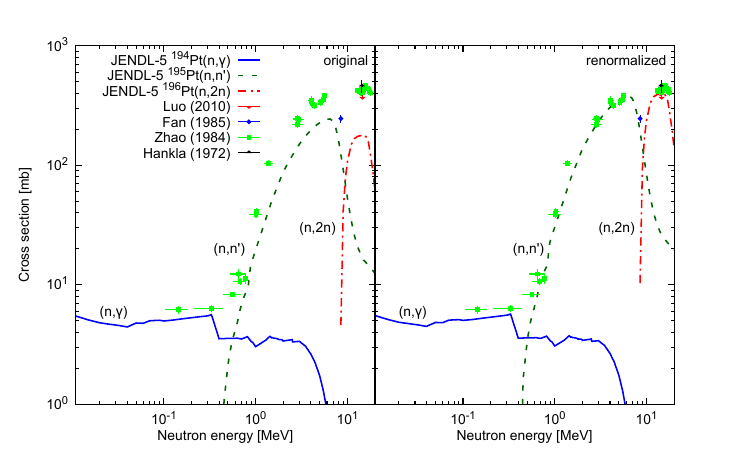"}
\caption{
$^{194}$Pt(n,$\gamma$)\mPt, $^{195}$Pt(n,n')\mPt and $^{196}$Pt(n,2n)\mPt cross sections in the JENDL-5/A library~\cite{Iwamoto2023} multiplied by the isotopic abundances of the target nuclides (left: original, right: renormalized) and the experimental cross sections~\cite{Luo2010,Fan1985,Zhao1984,Hankla1972}. 
}
\label{fig:J5}
\end{center}
\end{figure}

Figure~\ref{fig:J5} (left) shows the energy-dependent cross sections of these three reactions in the JENDL-5/A library weighted by the natural isotopic abundances of $^{194,195,196}$Pt along with 43 experimental data points of the \natPt(n,x)\mPt reaction~\cite{Luo2010,Fan1985,Zhao1984,Hankla1972} retrieved from the EXFOR library~\cite{Otuka2014}.
As shown in the figure, the JENDL-5/A cross sections particularly underestimate the experimental $^{195}$Pt(n,n’)\mPt and $^{196}$Pt(n,2n)\mPt cross sections.
Note that these JENDL-5/A cross sections were originally evaluated for the JENDL-5 library,
and its evaluation summary~\cite{Shibata2017} does not discuss the $^{195m}$Pt production cross sections.

To improve agreement with the experimental cross sections,
we renormalized the three JENDL-5/A cross sections by minimizing the differences from the measured cross sections using the least-squares method.
The relationship between the JENDL-5/A cross sections, renormalization factors and the experimental cross sections is expressed as
\begin{equation}
\mathbf{b}=X\mathbf{a}
\end{equation}
where $\mathbf{a}=(a_{194},a_{195},a_{196})^t$ is a vector of the renormalization factors for the three reactions,
$\mathbf{b}=\{b_i\} (i=1,\cdots,43)$ is a vector containing the 43 experimental cross sections,
and $X$ is a 43$\times$3 matrix whose columns are the JENDL-5/A $^{194}$Pt(n,$\gamma$)\mPt, $^{195}$Pt(n,n')\mPt, and $^{196}$Pt(n,2n)\mPt cross sections at the incident neutron energies of the experimental cross section $i$ multiplied by their natural isotopic abundances.
The optimal vector $\mathbf{a}$ minimizing the weighted residuals is obtained by solving the following least-squares equation:
\begin{eqnarray}
\mathbf{a}&=&(X^T V^{-1} X)^{-1} X^T V^{-1}\mathbf{b},
\label{eqn:lsf1}\\
M&=&(X^T V^{-1} X)^{-1},
\label{eqn:lsf2}
\end{eqnarray}
where $V$ is the diagonal matrix with the variances of the experimental cross sections,
and $M$ is the covariance matrix of $\mathbf{a}$.

Table~\ref{tab:renorm} summarizes the normalization factors determined by Eqs.~(\ref{eqn:lsf1}) and (\ref{eqn:lsf2}) for the $^{194}$Pt(n,$\gamma$)\mPt, $^{195}$Pt(n,n')\mPt and $^{196}$Pt(n,2n)\mPt cross sections in the JENDL-5/A library and their correlation coefficients.
Figure~\ref{fig:J5} (right) compares the renormalized JENDL-5/A cross sections with the experimental cross sections.
Compared with Fig.~\ref{fig:J5} (left), the renormalized JENDL-5/A cross sections show much better agreement with the experimental cross sections, indicating that the applied normalization effectively corrects the underestimations in the original JENDL-5/A library.

\begin{table}[hbtp]
\begin{center}
\caption{
Renormalization factors $a_i$ applied to JENDL-5/A $^x$Pt(n,x)\mPt cross sections obtained by the least-squares method.
}
\label{tab:renorm}
\begin{tabular}{llD{.}{.}{4}D{.}{.}{4}l}
\hline
$a_x$     &Factors         &\multicolumn{3}{l}{Correlation coefficients}\\
\hline
$a_{194}$ &1.01$\pm$0.05   &1       &       &                       \\
$a_{195}$ &1.56$\pm$0.03   &-0.229  &1      &                       \\
$a_{196}$ &2.24$\pm$0.02   &0.0259  &-0.119 &1                      \\
\hline
\end{tabular}
\end{center}
\end{table}

\subsection{Calculation of \mPt yields due to secondary neutron reactions}
\label{sec:yield}
The JENDL-5/A cross sections renormalized by us as described in Section~\ref{sec:leastsq} were interpolated and averaged over the neutron energy group defined in Section~\ref{sec:nfield}.
By using the group-wise cross sections, the number of \mPt nuclei produced by secondary neutrons per incident alpha particle in each platinum foil was calculated by
\begin{equation}
Y=\sum_i \sigma_i \phi_i\, \rho V,
\label{eqn:yield}
\end{equation}
where the summation is taken over the neutron energy group $i$, and the parameters are defined as follows:

\noindent
\begin{tabular}{ll}
$\sigma_i$: &Renormalized JENDL-5/A \natPt(n,x)\mPt production cross section for the energy group $i$ [cm$^2$] \\
$\phi_i$:   &Neutron fluence in the energy group $i$ [neutrons/cm$^2$/$\alpha$] \\
$\rho$:     &Volume number density of Pt nuclei [nuclei/cm$^3$] \\
$V$ :       &Volume of the platinum foil [cm$^3$] \\
\end{tabular}

\noindent
Note that the upper energy boundary of the JENDL-5/A library is 20~MeV and we set $\sigma_i$=0 above 20~MeV.
The PHITS calculation shows a small fraction of neutrons above 20~MeV and we will discuss its influence in Section~\ref{sec:highE}.

\section{Results}
\label{sec:results}

\subsection{Neutron energy spectra in the stacked foils}
\label{sec:nspect}
\begin{figure}[hbtp]
\begin{center}
\includegraphics[width=0.8\textwidth]{"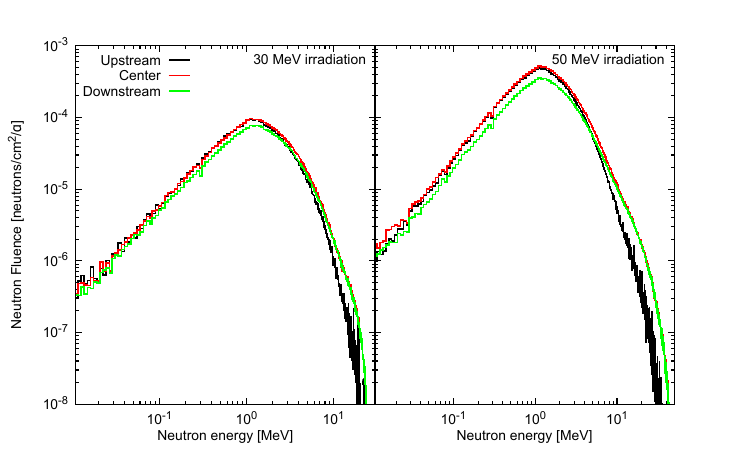"}
\caption{
Neutron energy spectra calculated with the INC option in the 1st (Upstream), 5th (Center) and 9th (Downstream) platinum foils from the upstream side under the 30~MeV alpha-particle irradiation (left), and in the 1st (Upstream), 10th (Center) and 19th (Downstream) platinum foils from the upstream side under the 50~MeV alpha-particle irradiation (right).
Each ten neutron energy groups introduced in Section~\ref{sec:nfield} collapse to one bin for better visual clarity. 
}
\label{fig:nfield}
\end{center}
\end{figure}

Figure~\ref{fig:nfield} (left) presents the neutron energy spectra calculated with the INC option for the case of the 30~MeV alpha-particle irradiation.
The figure shows the results for the 1st, 5th, and 9th platinum foils from the upstream side,
corresponding to the mean alpha-particle energies\footnote{
The mean alpha-particle energy is the mean of the alpha-particle energies at the upstream and downstream sides of the foil estimated with the stopping power of the foil materials.}
of 28.4~MeV, 19.5~MeV, and 5.7~MeV, respectively.
A prominent peak around 1~MeV is observed in all three cases,
indicating a large number of neutrons in this energy region.
Similar spectral features were seen in the other foils as well as the spectra calculated with the NDL option.

Figure~\ref{fig:nfield} (right) shows the corresponding neutron energy spectra for the 50~MeV alpha-particle irradiation,
again using the INC option.
The results for the 1st, 10th, and 19th platinum foils from the upstream side are displayed. In the 1st and 10th foils, the mean alpha-particle energies are 50.2~MeV and 28.8~MeV, respectively,
while the alpha particles did not reach the 19th foil.
As in the 30~MeV irradiation, the fluence peaks around 1~MeV, and similar trends were observed across all foils. This tendency was also reproduced in the calculation with the NDL option.

In both 30~MeV and 50~MeV cases, the neutron fluence near the peak energy region ($\sim$1~MeV) decreases in the downstream foils,
whereas in the high-energy tail region,
the upstream foils showed lower fluences.
To investigate this behavior,
we performed additional simulations with the INC option excluding the compound nuclear process implemented by Generalized Evaporation Model (GEM)~\cite{Furihata2000}. 

\begin{figure}[hbtp]
\begin{center}
\includegraphics[width=0.8\textwidth]{"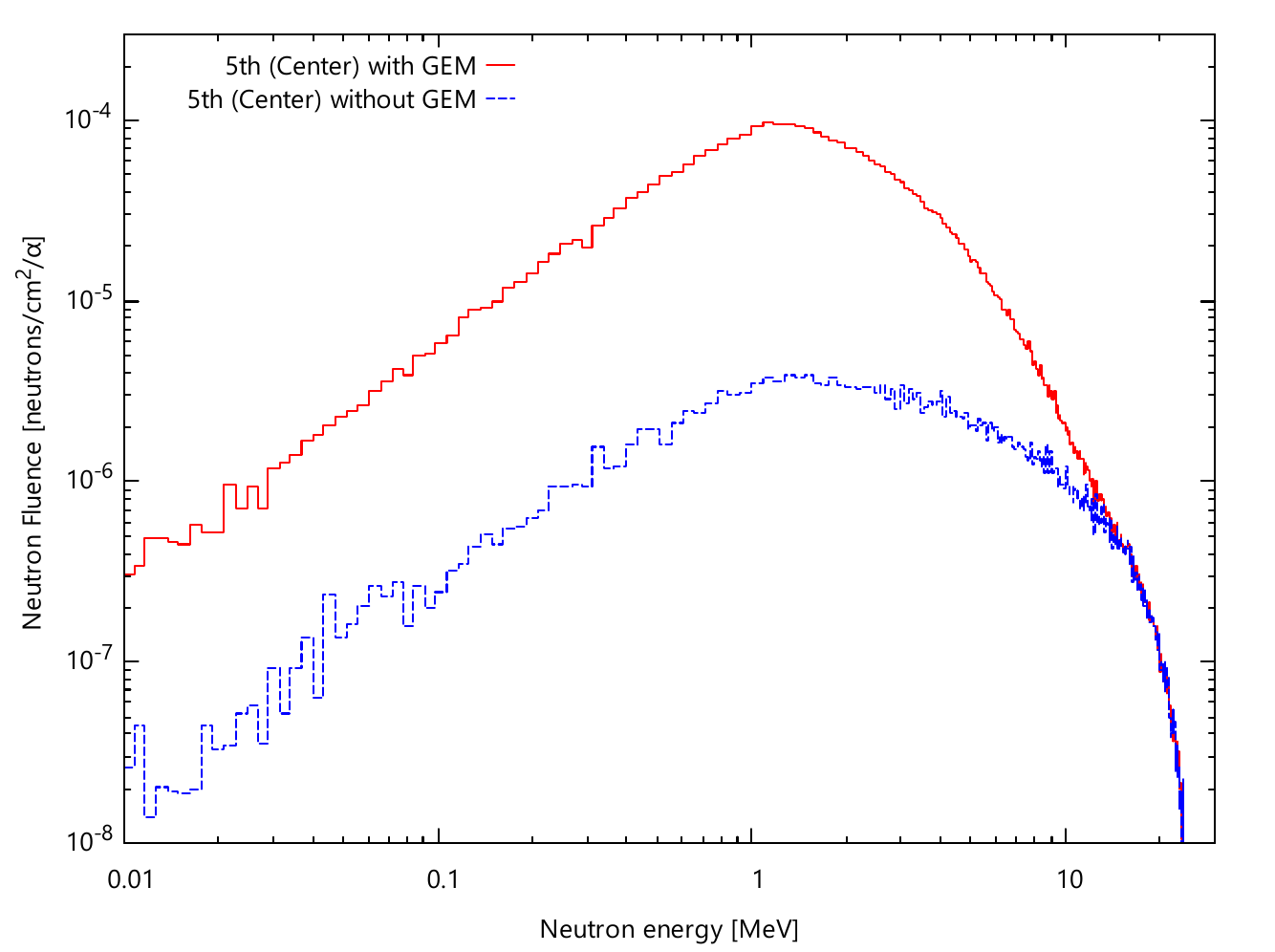"}
\caption{
Comparison of neutron energy spectra with and without GEM in the 5th platinum foil from the upstream side under the 30~MeV alpha-particle irradiation calculated with the INC option.
Each ten neutron energy groups introduced in Section~\ref{sec:nfield} collapse to one bin for better visual clarity. 
}
\label{fig:GEM}
\end{center}
\end{figure}

Figure~\ref{fig:GEM} compares the neutron energy spectra with and without inclusion of the evaporation process implemented by GEM in the 5th foil from the upstream side under the 30~MeV alpha-particle irradiation in the simulation.
This comparison reveals that the neutrons near the peak energy are predominantly produced via the compound nucleus process,
which tends to emit neutrons isotropically.
In contrast,
the high-energy neutrons are mainly generated through intranuclear cascade process, which have forward angular distributions.
Therefore,
the reduction of the neutron fluence at the peak energy observed in the downstream foils is attributed to isotropic production and emission of the neutrons in the peak region through the compound nucleus process,
and the yield of such neutrons decreases  for alpha particles with lower energies.
On the other hand,
the lower neutron fluence at the high energy tail in the upstream foils is likely because high-energy neutrons are mainly generated through the intranuclear cascade process,
which preferentially emits neutrons in the forward (downstream) direction.
As a result,
more high energy neutrons reach the downstream foils,
while their fluence is reduced in the upstream foils.

\subsection{Influence of secondary neutrons on measured \natPt(\a,x)\mPt cross section}
\label{sec:influence}

The contribution of the secondary neutrons to the measured \natPt(\a,x)\mPt cross section (mb) is calculated by
\begin{equation}
\sigma_n=Y/n
\end{equation}
where $Y$ is the \mPt yield produced by the secondary neutrons in the foil (nuclei/\a) determined by Eq.~(\ref{eqn:yield}),
and $n$ is the areal number density of the Pt nuclei in the foil (nuclei/mb). The uncertainty in $\sigma_n$ may be propagated from the standard deviation in $Y$ (less than 1\%) due to random sampling in the PHITS calculation.

\begin{figure}[hbtp]
\begin{center}
\includegraphics[width=0.8\textwidth]{"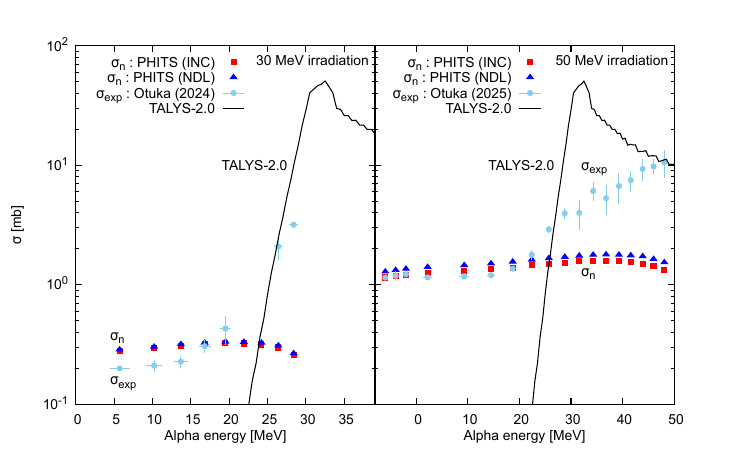"}
\caption{
\natPt(\a,x)\mPt cross sections under 30~MeV (left) and 50~MeV (right) alpha-particle irradiation.
The data points shown at -2, -4 and -6 MeV on the right panel are for the 17th, 18th, and 19th platinum foils,
which are not exposed to alpha-particles.
}
\label{fig:excfun}
\end{center}
\end{figure}

Figures~\ref{fig:excfun} left and right present $\sigma_n$ under the 30~MeV and 50~MeV alpha-particle irradiation, respectively.
The 17th, 18th, and 19th platinum foils from the upstream side under the 50~MeV alpha-particle irradiation were not exposed to alpha particles,
namely any activation observed in these foils is attributed solely to secondary neutrons.
The estimated $\sigma_n$ of these foils are plotted at -2, -4, and -6~MeV, respectively, in Fig.~\ref{fig:excfun} (right) although these are not actual incident alpha-particle energies.
It should be noted that production of \mPt in alpha-induced reactions such as $^{195}$Pt(\a,\a')\mPt is not included in the present PHITS calculations.

As shown in these figures, $\sigma_n$ is comparable with the measured \natPt(\a,x)\mPt cross section $\sigma_\mathrm{exp}$ in the low-energy region below the Coulomb barrier ($\sim$23~MeV).
This indicates that the \mPt production in this energy region is primarily due to presence of the secondary neutrons.

For the foils irradiated by alpha particles at very low energy,
$\sigma_\mathrm{exp}$ is exhausted by $\sigma_n$,
and calculated $\sigma_n$ is even higher than $\sigma_\mathrm{exp}$.
This overestimation is maximum 40\% and 20\% for the 30 and 50~MeV irradiation, respectively.
It means $\sigma_\mathrm{exp}-\sigma_n$ becomes negative,
and we cannot use $\sigma_n$ directly for correction of $\sigma_\mathrm{exp}$ for the secondary neutron contribution. 

\section{Discussion}
\label{sec:discussion}

\subsection{Validation and practical correction of secondary neutron contribution}
\label{sec:correction}
For the 50~MeV irradiation case,
the best agreement (within few percent) between $\sigma_n$ and $\sigma_\mathrm{exp}$ is achieved for the last three Pt foils not irradiated by alpha particles.
Figure~\ref{fig:excfun} also shows that $\sigma_n$ has weak foil dependence.
It would be therefore good practice for correction of the secondary neutron effect by adding some foils at the end of the target stack longer than the range of the initial alpha-particle beam energy.
The cross sections measured for these extra foils as $\sigma_n$ can be adopted for subtraction from the cross sections measured in entire foils as long as the neutron fluence does not show strong foil dependence.

Tak\'{a}cs et al.~\cite{Takacs2023} adopt a similar magnitude of the background cross section ($\sim$1.2~mb) for the $^{181}$Ta(\a,x)$^{182g}$Ta cross sections measured by irradiation of a foil stack consisting of a series of Ta-Ti-Ti by a 50~MeV alpha-particle beam.
They consider that the $^{181}$Ta(n,$\gamma$)$^{182g}$Ta contribution is due to thermal neutrons.
It would be interesting to examine if the procedure developed by the present work works reasonably for a $^{181}$Ta(\a,x)$^{182g}$Ta cross section measurement,
to which secondary neutrons can contribute only by the $^{181}$Ta(n,$\gamma$)$^{182g}$Ta neutron capture process.

\subsection{Impact of neutrons above 20~MeV on \mPt production}
\label{sec:highE}
As mentioned in Section~\ref{sec:yield},
the incident neutron energy of the JENDL-5/A library is limited up to 20~MeV,
and we assumed that the contribution of the neutrons above 20~MeV to \mPt production is negligible.
To evaluate the impact of this omission,
$\sigma_n$ was estimated by replacing the JENDL-5/A library with the TENDL-2023 library for $\sigma_i$ in Eq.~(\ref{eqn:yield}) as the latter provides the neutron-induced \mPt production cross sections up to 200~MeV.
All energetically possible \mPt production routes in the neutron induced reaction on natural platinum were considered: $^{194}$Pt(n,$\gamma$)\mPt, $^{195}$Pt(n,n')\mPt, $^{196}$Pt(n,2n)\mPt, and $^{198}$Pt(n,4n)\mPt.

\begin{figure}[hbtp]
\begin{center}
\includegraphics[width=0.8\textwidth]{"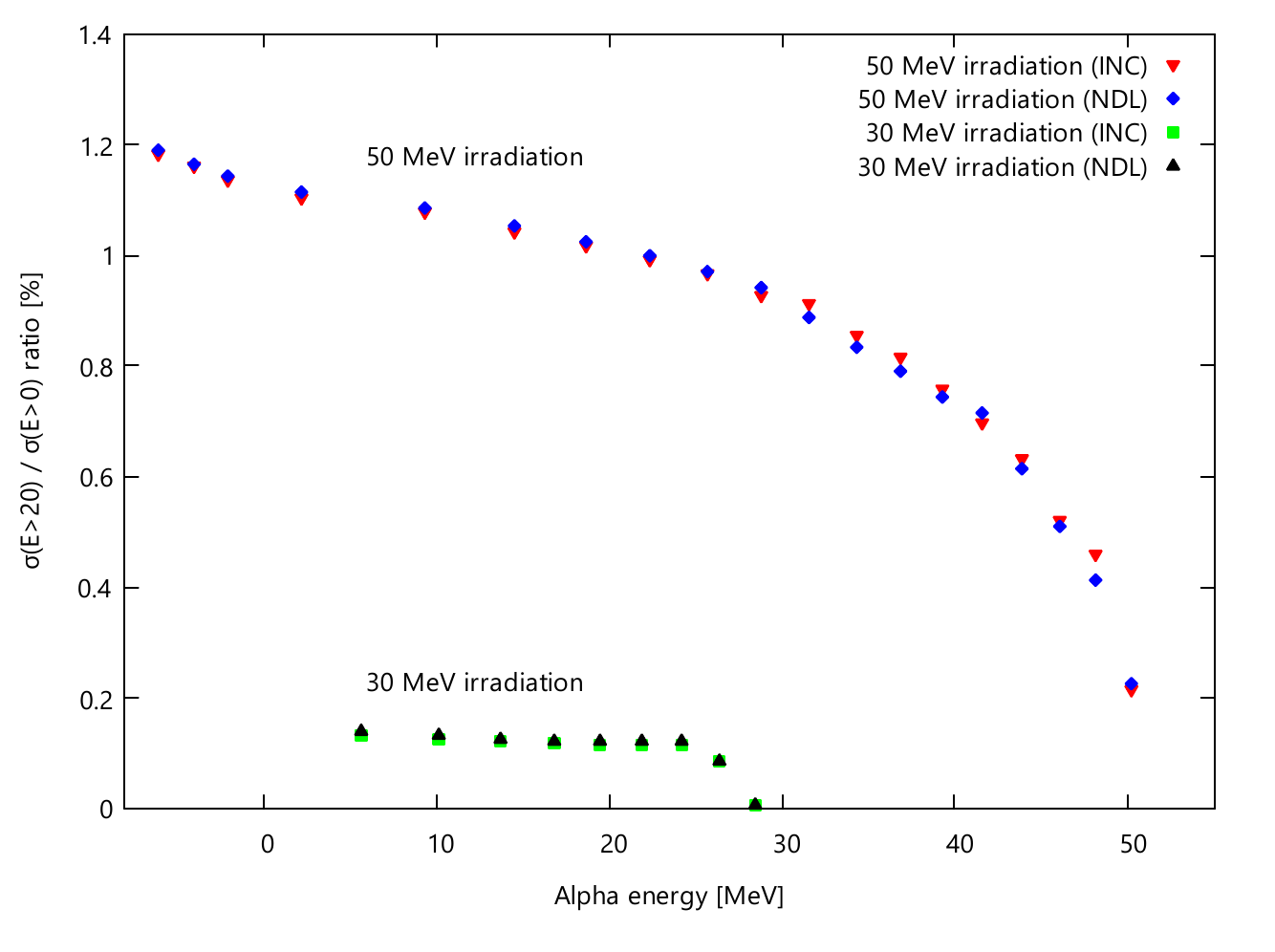"}
\caption{
Contribution of high energy ($E>$20~MeV) neutrons to \mPt production using TENDL-2023 cross sections under 30 and 50~MeV alpha-particle irradiation.
The data points shown at -2, -4 and -6 MeV are for the 17th, 18th, and 19th platinum foils,
which are not exposed to alpha-particles.
CP Total denotes the sum of contributions from protons, deuterons, tritons, alpha particles and $^3$He particles.
}
\label{fig:impact}
\end{center}
\end{figure}

Figure~\ref{fig:impact} shows the ratio of the \mPt production cross section originating from neutrons with energies above 20~MeV to that originating from the entire neutron energy range calculated using the \natPt(n,x)\mPt cross sections in the TENDL-2023 library for both 30~MeV and 50~MeV alpha-particle irradiations.
As in Figure~\ref{fig:excfun} (right),
the results for the 17th, 18th, and 19th platinum foils from the upstream side under the 50~MeV alpha-particle irradiation are plotted at -2, -4, and -6~MeV, respectively.
The contribution from the neutrons above 20~MeV was found to be at most $\sim$0.1\% for 30~MeV irradiation,
and $\sim$1\% for 50~MeV irradiation,
which is within the accuracy of the nuclear data libraries.

These results suggest that the use of JENDL-5/A cross sections, limited to 0--20~MeV neutron energies,
is sufficient for evaluating $\sigma_n$ under the experimental conditions considered in this study.

\subsection{Impact of secondary charged particles on \mPt production}
\label{sec:secchar}
The Hannover group~\cite{Michel1995} discusses the secondary particle effect not only for secondary neutrons but also for secondary protons for their high energy proton activation.
For completeness,
we also evaluated the contribution of secondary charged particles to the production of \mPt by calculating the fluences of protons, deuterons, tritons, alpha particles, and $^3$He particles at each platinum foil by using PHITS with the INC option.
Based on these fluences and the \mPt production cross sections in the TENDL-2023 library,
the yield of \mPt produced by secondary charged particles was estimated.
The considered \mPt production reactions are the \natPt(p,x)\mPt, \natPt(d,x)\mPt, \natPt(t,x)\mPt, \natPt(\a,x)\mPt and \natPt($^3$He,x)\mPt reactions.

\begin{figure}[hbtp]
\begin{center}
\includegraphics[width=0.8\textwidth]{"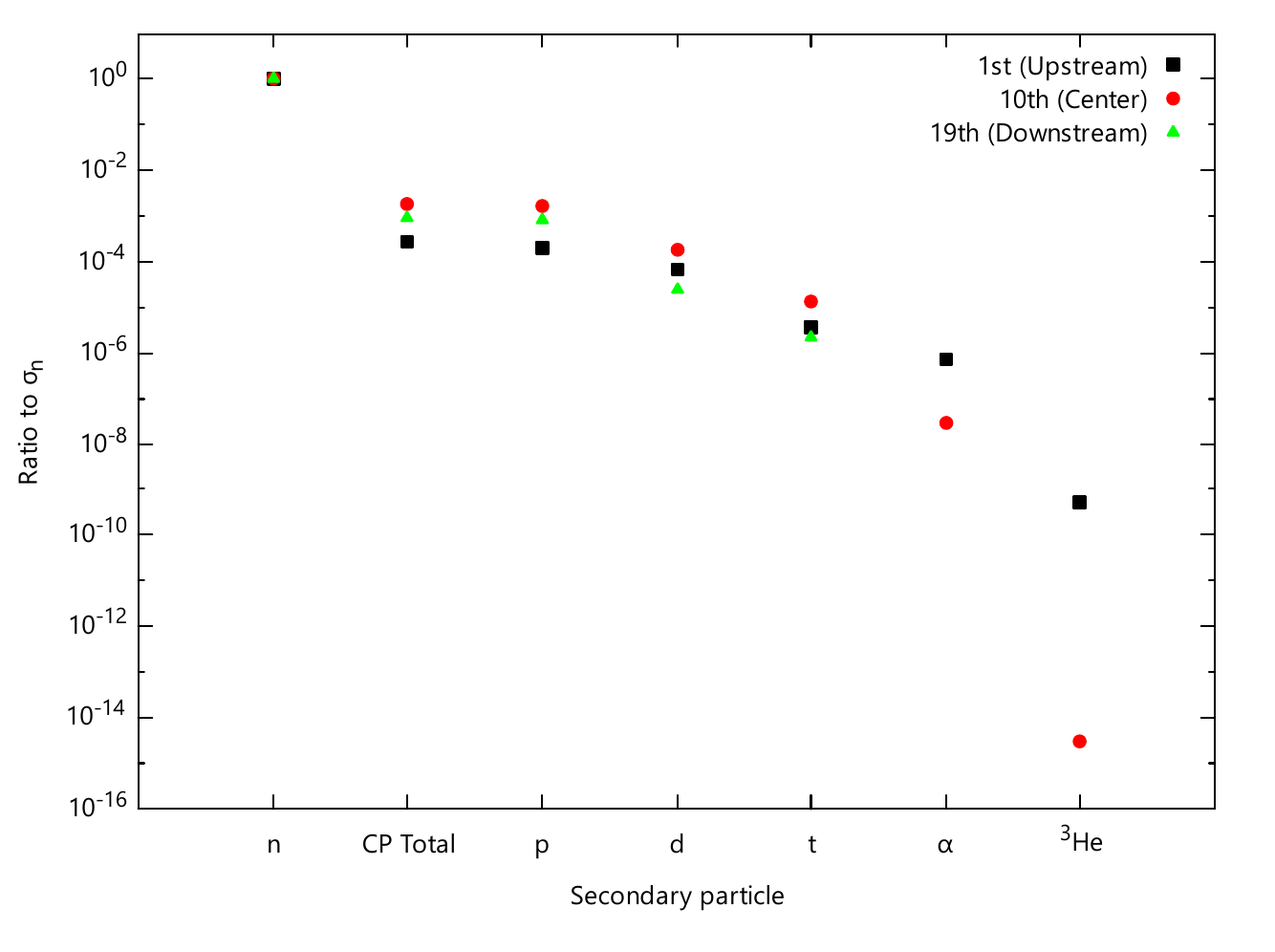"}
\caption{
Ratios of \mPt production from secondary charged particles to that from secondary neutrons in the 1st (Upstream), 10th (Center), and 19th (Downstream) platinum foils under 50~MeV alpha-particle irradiation. 
}
\label{fig:secchar}
\end{center}
\end{figure}

Figure~\ref{fig:secchar} presents the ratios of \mPt production induced by a secondary charged particle to that induced by a secondary neutron for the 1st, 10th and 19th platinum foils from the upstream side under 50~MeV alpha-particle irradiation.
CP Total on the figure denotes the sum of contributions from protons, deuterons, tritons, alpha particles and $^3$He particles.
To exclude contribution of the primary alpha-particles to the \natPt(\a,x)\mPt events, the contributions of the alpha particles with their energies close to the primary ones (above 49.45~MeV for the 1st foil and 27.80~MeV for the 10th foil) were not taken into account.
Such exclusion is not necessary for the 19th foil since it is not irradiated by the primary alpha particles.
In the 19th foil, no data points appear for alpha particles and $^3$He in Figure~\ref{fig:secchar} since their production cross sections were calculated to be zero.

As shown in the figure, the \mPt production via secondary charged particles is less than 0.2\% of that via secondary neutrons in all investigated foils, indicating that their contribution is negligible under the present experimental conditions.

\section{Summary}
The contribution of the secondary neutrons to the \natPt(\a,x)\mPt cross section measured by the stacked foil activation method was studied in the subbarrier region.
First,
the energy spectra of the secondary neutrons produced by interaction of alpha-particles with the stacked platinum foils were calculated using the particle transport simulation code PHITS with the INCL model and TENDL-2023 nuclear data library.
Then, the JENDL-5/A cross sections for the $^{194}$Pt(n,$\gamma$)\mPt, $^{195}$Pt(n,n')\mPt, and $^{196}$Pt(n,2n)\mPt reactions were renormalized through least-squares fit to the experimental cross sections.
Finally, the contribution of the secondary neutrons to the measured \natPt(\a,x)\mPt production cross sections were calculated using the obtained neutron spectra and the renormalized JENDL-5/A cross sections,
and the results were compared with the experimental \natPt(\a,x)\mPt cross sections measured by us previously.

The results showed good agreement between the calculated and experimental \mPt production cross sections below the Coulomb barrier,
suggesting that the \mPt production observed in this region is primarily attributed to the secondary neutrons.
These findings highlight the importance of considering the influence of secondary neutrons when interpreting charged-particle activation cross section measurements,
particularly in the low-energy region below the Coulomb barrier.
We also analyzed the influence of secondary charged particles,
and concluded that their influence is negligible in comparison with the influence of the secondary neutrons.

The present work demonstrated that the neutron field characterized by the particle transport simulation combined with the renormalized JENDL-5/A cross sections explains the $^{195m}$Pt yields.
However,
we are not quite sure to what extent our simulation reproduces the actual neutron field.
Apart from the thick target neutron spectra for bismuth irradiated by alpha-particles at 30 MeV~\cite{Sugihara2020},
we could not find a similar spectrum measurement usable for benchmarking.
It would be worthwhile to assess the neutron field experimentally by attaching neutron dosimetry foils near the beamline during irradiation and to check if the activities of the dosimetry foils can be reasonably described by the neutron field predicted by simulation.

The \natPt(n,x)\mPt elemental cross section derived from the JENDL-5/A $^{194,195,196}$Pt(n,x)\mPt isotopic cross sections required us renormalization for better description of the experimental elemental cross sections.
Validation of the evaluated isotopic cross sections against the experimental elemental cross section could be time-consuming for the evaluators.
Nevertheless, we would encourage the evaluators to enhance usability and reliability of the library by systematic adjustment of the evaluated cross section with the experimental elemental cross sections.

\begin{acknowledgement}
We thank Konno Chikara and Tatsuhiko Ogawa (JAEA) for discussion on determination of yields of radionuclides produced by the secondary neutrons by the PHITS code, Nobuyuki Iwamoto (JAEA) for discussion on the contents of the JENDL-5/A library, and Arjan Koning (IAEA) for discussion on the mechanism of neutron production in the TALYS code.
We received valuable comments on our result from S\'{a}ndor Tak\'{a}cs (ATOMKI).
Production of the ACE files by NJOY2016 was performed under instruction by Oscar Cabellos (Universidad Polit\'{e}cnica Madrid).
One of us (Y.F.) performed this work partly under the internship program of International Atomic Energy Agency. 
\end{acknowledgement}

\nocite{*}
\bibliography{arxiv}

\begin{thebibliography}{10}

\bibitem{Schillebeeckx2012}
P.~Schillebeeckx, B.~Becker, Y.~Danon, K.~Guber, H.~Harada, J.~Heyse, A.~R.
  Junghans, S.~Kopecky, C.~Massimi, M.~C. Moxon, N.~Otuka, I.~Sirakov, and
  K.~Volev.
\newblock Determination of resonance parameters and their covariances from
  neutron induced reaction cross section data.
\newblock {\em Nuclear Data Sheets}, 113:3054--3100, 2012.

\bibitem{Clark1982}
J.~L. Clark, P.~E. Haustein, T.~J. Ruth, J.~Hudis, and A.~A. Caretto.
\newblock Inclusive $\pi^-$ production with 200~{MeV} protons: {R}adiochemical
  study of the $^{209}${Bi}(p,$\pi^-xn$) $^{210-x}${At} reactions.
\newblock {\em Physical Review C}, 26:2073--2083, 1982.

\bibitem{Michel1995}
R.~Michel, M.~Gloris, H.~J. Lange, I.~Leya, M.~L\"{u}pke, U.~Herpers,
  B.~Dittrich-Hannen, R.~R\"{o}sel, Th. Schiekel, D.~Filges, P.~Dragovitsch,
  M.~Suter, H.~J. Hofmann, W.~W\"{o}lfli, P.~W. Kubik, H.~Baur, and R.~Wieler.
\newblock Nuclide production by proton-induced reactions on elements (6
  \ensuremath{\leq} {Z} \ensuremath{\leq} 29) in the energy range from 800 to
  2600~{MeV}.
\newblock {\em Nuclear Instruments and Methods in Physics Research Section B:
  Beam Interactions with Materials and Atoms}, 103:183--222, 1995.

\bibitem{Luepke1993}
M~L\"{u}pke.
\newblock {\em Untersuchung zur Wechselwirkung galaktischer Protonen mit
  Meteoroiden: Dicktarget-Simulationsexperimente und Messung von
  D\"{u}nntarget-Wirkungsquerschnitten}.
\newblock PhD thesis, Universit\"{a}t Hannover, 1993.

\bibitem{Armstrong1972}
T.~W. Armstrong and K.~C. Chandler.
\newblock {HETC}: {A} high energy transport code.
\newblock {\em Nuclear Science and Engineering}, 49:110--111, 1972.

\bibitem{Khandaker2013}
Mayeen~Uddin Khandaker, Hiromitsu Haba, Jumpei Kanaya, and Naohiko Otuka.
\newblock Excitation functions of (d,x) nuclear reactions on natural titanium
  up to 24~{MeV}.
\newblock {\em Nuclear Instruments and Methods in Physics Research Section B:
  Beam Interactions with Materials and Atoms}, 296:14--21, 2013.

\bibitem{Otuka2024}
Naohiko Otuka, S{\'a}ndor Tak{\'a}cs, Masayuki Aikawa, Shuichiro Ebata, and
  Hiromitsu Haba.
\newblock Isomer production studied with simultaneous decay curve analysis for
  alpha-particle induced reactions on natural platinum up to 29~{MeV}.
\newblock {\em The European Physical Journal A}, 60:195, 2024.

\bibitem{Otuka2025}
Naohiko Otuka, Masayuki Aikawa, S{\'a}ndor Tak{\'a}cs, Damdinsuren Gantumur,
  Shuichiro Ebata, Lkhagvasuren Bold, Akihiro Nambu, and Hiromitsu Haba.
\newblock Energy dependence of isomeric ratios for alpha-particle-induced
  reactions on natural platinum up to 50~{MeV} studied by simultaneous decay
  curve analysis.
\newblock {\em The European Physical Journal A}, 61:184, 2025.

\bibitem{Sugihara2020}
K.~Sugihara, N.~Shigyo, E.~Lee, T.~Sanami, and K.~Tanaka.
\newblock Measurement of thick target neutron yields from 7~{MeV}/u
  \ensuremath{\alpha} incidence on $^{209}${Bi}.
\newblock {\em Nuclear Instruments and Methods in Physics Research Section B:
  Beam Interactions with Materials and Atoms}, 470:15--20, 2020.

\bibitem{Boudard2013}
A.~Boudard, J.~Cugnon, J.~C. David, S.~Leray, and D.~Mancusi.
\newblock New potentialities of the {Li\`ege} intranuclear cascade model for
  reactions induced by nucleons and light charged particles.
\newblock {\em Physical Review C}, 87:014606, 2013.

\bibitem{Sato2018}
Tatsuhiko Sato, Yosuke Iwamoto, Shintaro Hashimoto, Tatsuhiko Ogawa, Takuya
  Furuta, Shin-Ichiro Abe, Takeshi Kai, Pi-En Tsai, Norihiro Matsuda, Hiroshi
  Iwase, Nobuhiro Shigyo, Lembit Sihver, and Koji Niita.
\newblock Features of particle and heavy ion transport code system ({PHITS})
  version 3.02.
\newblock {\em Journal of Nuclear Science and Technology}, 55:684--690, 2018.

\bibitem{Iwamoto2023}
Osamu Iwamoto, Nobuyuki Iwamoto, Satoshi Kunieda, Futoshi Minato, Shinsuke
  Nakayama, Yutaka Abe, Kohsuke Tsubakihara, Shin Okumura, Chikako Ishizuka,
  Tadashi Yoshida, Satoshi Chiba, Naohiko Otuka, Jean-Christophe Sublet, Hiroki
  Iwamoto, Kazuyoshi Yamamoto, Yasunobu Nagaya, Kenichi Tada, Chikara Konno,
  Norihiro Matsuda, Kenji Yokoyama, Hiroshi Taninaka, Akito Oizumi, Masahiro
  Fukushima, Shoichiro Okita, Go~Chiba, Satoshi Sato, Masayuki Ohta, and Saerom
  Kwon.
\newblock Japanese evaluated nuclear data library version 5: {JENDL}-5.
\newblock {\em Journal of Nuclear Science and Technology}, 60:1--60, 2023.

\bibitem{Ogawa2025}
Tatsuhiko Ogawa.
\newblock Prediction of composite neutron source spectra by combination of
  {JENDL-5} and {PHITS}.
\newblock {\em Annals of Nuclear Energy}, 216:111256, 2025.

\bibitem{Iwamoto2017}
Yosuke Iwamoto, Tatsuhiko Sato, Shintaro Hashimoto, Tatsuhiko Ogawa, Takuya
  Furuta, Shin-Ichiro Abe, Takeshi Kai, Norihiro Matsuda, Ryuji Hosoyamada, and
  Koji Niita.
\newblock Benchmark study of the recent version of the {PHITS} code.
\newblock {\em Journal of Nuclear Science and Technology}, 54:617--635, 2017.

\bibitem{Nara1999}
Y.~Nara, N.~Otuka, A.~Ohnishi, K.~Niita, and S.~Chiba.
\newblock Relativistic nuclear collisions at 10$a$ {GeV} energies from {p+Be}
  to {Au+Au} with the hadronic cascade model.
\newblock {\em Physical Review C}, 61:024901, 1999.

\bibitem{Hayakawa1955}
Satio Hayakawa, Mitsuji Kawai, and Ken Kikuchi.
\newblock Nuclear reactions at moderate energies and {Fermi} gas model.
\newblock {\em Progress of Theoretical Physics}, 13:415--441, 1955.

\bibitem{David2015}
J.~C. David.
\newblock Spallation reactions: {A} successful interplay between modeling and
  applications.
\newblock {\em The European Physical Journal A}, 51:68, 2015.

\bibitem{Koning2023}
Arjan Koning, Stephane Hilaire, and Stephane Goriely.
\newblock {TALYS}: modeling of nuclear reactions.
\newblock {\em The European Physical Journal A}, 59:131, 2023.

\bibitem{Koning2019}
A.~J. Koning, D.~Rochman, J.~Ch. Sublet, N.~Dzysiuk, M.~Fleming, and S.~Van
  Der~Marck.
\newblock {TENDL}: Complete nuclear data library for innovative nuclear science
  and technology.
\newblock {\em Nuclear Data Sheets}, 155:1--55, 2019.

\bibitem{MacFarlane2010}
R.~E. MacFarlane and A.~C. Kahler.
\newblock Methods for processing {ENDF/B-VII} with {NJOY}.
\newblock {\em Nuclear Data Sheets}, 111:2739--2890, 2010.

\bibitem{Sartori1991}
E~Sartori.
\newblock {VITAMIN-J}, a standard group cross section library structure for
  reactor shielding and fusion neutronics applications.
\newblock Technical Report JEF/DOC-315 Revision 4 - Draft, OECD/NEA Data Bank,
  1991.

\bibitem{Iwamoto2022}
N~Iwamoto.
\newblock Summary of {JENDL-5} activation cross section data.
\newblock In {\em Proceedings of the 2022 Fall Meeting of Atomic Energy Society
  of Japan}, page 3N\_PL01, 2022.

\bibitem{Luo2010}
Junhua Luo, Xinxing Wang, Zhenlai Liu, Fei Tuo, and Xiangzhong Kong.
\newblock Cross-sections for formation of $^{195m}${Pt} through
  $^\mathrm{nat}${Pt}(n,x)$^{195m}${Pt} reaction over neutron energy range
  13--15~{MeV}.
\newblock {\em Radiation Physics and Chemistry}, 79:1018--1021, 2010.
\newblock {EXFOR} 32674.

\bibitem{Fan1985}
Peiguo Fan, Wenrong Zhao, Dan Teng, and Hanlin Lu.
\newblock Measurements of cross sections for some reactions induced by
  8.62~{MeV} neutrons.
\newblock {\em Chinese Journal of Nuclear Physics}, 7:242--245, 1985.
\newblock {EXFOR} 30733.

\bibitem{Zhao1984}
Wenrong Zhao, Hanlin Lu, Peiguo Fan, Baosheng Yu, Enchen Zhou, and Dan Teng.
\newblock Measurement of cross section for reaction {Pt}(n,x)$^{195m}${Pt}.
\newblock {\em Chinese Journal of Nuclear Physics}, 6:342--346, 1984.
\newblock {EXFOR} 30715.

\bibitem{Hankla1972}
A.~K. Hankla, R.~W. Fink, and J.~H. Hamilton.
\newblock Neutron activation cross sections at 14.4 {MeV} for some naturally
  occurring heavy elements in the region 76$\le{Z}\le$82.
\newblock {\em Nuclear Physics A}, 180:157--176, 1972.
\newblock {EXFOR} 10244.

\bibitem{Otuka2014}
N.~Otuka, E.~Dupont, V.~Semkova, B.~Pritychenko, A.~I. Blokhin, M.~Aikawa,
  S.~Babykina, M.~Bossant, G.~Chen, S.~Dunaeva, R.~A. Forrest, T.~Fukahori,
  N.~Furutachi, S.~Ganesan, Z.~Ge, O.~O. Gritzay, M.~Herman, S.~Hlava\v{c},
  K.~Kat\={o}, B.~Lalremruata, Y.~O. Lee, A.~Makinaga, K.~Matsumoto,
  M.~Mikhaylyukova, G.~Pikulina, V.~G. Pronyaev, A.~Saxena, O.~Schwerer, S.~P.
  Simakov, N.~Soppera, R.~Suzuki, S.~Tak{\'a}cs, X.~Tao, S.~Taova,
  F.~T{\'a}rk{\'a}nyi, V.~V. Varlamov, J.~Wang, S.~C. Yang, V.~Zerkin, and
  Y.~Zhuang.
\newblock Towards a more complete and accurate experimental nuclear reaction
  data library ({EXFOR}): International collaboration between nuclear reaction
  data centres ({NRDC}).
\newblock {\em Nuclear Data Sheets}, 120:272--276, 2014.

\bibitem{Shibata2017}
Keiichi Shibata.
\newblock Evaluation of neutron nuclear data on platinum isotopes.
\newblock {\em Journal of Nuclear Science and Technology}, 54:147--157, 2017.

\bibitem{Furihata2000}
S.~Furihata.
\newblock Statistical analysis of light fragment production from medium energy
  proton-induced reactions.
\newblock {\em Nuclear Instruments and Methods in Physics Research Section B:
  Beam Interactions with Materials and Atoms}, 171:251--258, 2000.

\bibitem{Takacs2023}
S.~Tak\'{a}cs, F.~Ditr\'{o}i, Z.~Sz\H{u}cs, M.~Aikawa, H.~Haba, Y.~Toyoeda,
  G.~Damdinsuren, and S.~Ebata.
\newblock Activation cross section measurement of alpha-particle induced
  nuclear reactions on tantalum.
\newblock {\em Nuclear Instruments and Methods in Physics Research Section B:
  Beam Interactions with Materials and Atoms}, 545:165127, 2023.

\end{thebibliography}

\end{document}